\documentstyle[12pt,psfig,epsfig]{article}
\begin{document}
\pagestyle{empty}
\oddsidemargin -0.4cm
\topmargin -1.25cm
\begin{flushright}
hep-ph/0202045
\end{flushright}
\setcounter{footnote}{0}
\def \eett {$e^+ e^- \ra t \o t$}
\def \ggtt {$\g \g \ra t \o t$}
\def \o {\overline}
\def \ttbar{$t \overline{t}~$}
\def \tt{t \overline{t}}
\def \eebar{$e^+e^-$}
\def \ee{e^+e^-}
\def \gg{\gamma\gamma}
\def \ra{\rightarrow}
\def \beq{\begin{equation}}
\def \eeq{\end{equation}}
\def \bea{\begin{eqnarray}}
\def \eea{\end{eqnarray}}
\def \f{\frac}
\def \ee {e^+e^-}
\def \ra {\rightarrow}
\def \g {\gamma}
\def \cvg {c_v^{\gamma}}
\def \cag {c_a^{\gamma}}
\def \cvz {c_v^Z}
\def \caz {c_a^Z}
\def \cdg {c_d^{\gamma}}
\def \cdz {c_d^Z}
\def \cdgz {c_d^{\gamma,Z}}
\def \t{$t\;$}
\def \toverline{$\overline t$}
\def \tbar{$\overline t\;$}
\def \el{E_l}
\def \ttt{\theta_t}
\def \tht{\theta_t}
\def \thl{\theta_l}
\def \thetal{\theta_l}       
\begin{center}
{\large \boldmath \bf CP violation in open \ttbar production at a 
linear collider
\footnote{Talk presented at the 4th ACFA Workshop on 
Physics/Detector at the 
Linear Collider, Beijing, October 31 - November 2, 2001}
}
\vskip 0.4cm
Saurabh D. Rindani
\vskip .1cm
{\it Theory Group,
Physical Research Laboratory \\
Navrangpura,
Ahmedabad 380 009,
India}
\end{center}
\vskip .4cm
\centerline{\small \bf Abstract}
\vskip .1cm
\begin{quote}
{  \small
CP-violating asymmetries in the processes \eett~ and \ggtt~ can provide information
about CP-violating couplings of the top quark, whose presence would signal physics 
beyond the standard model. Work on studies of different asymmetries has been described.
Special emphasis has been laid on asymmetries in charged lepton distributions arising
from top decay. The effect of longitudinal beam polarization has been discussed.
Limits possible on the electric dipole moment of the top quark from different angular 
asymmetries have been obtained.     
}
\end{quote}

\vskip 0.4cm
Linear colliders can provide a clean environment for the study of CP 
violation in top-quark couplings in the process $\ee \ra \tt$ and also in
$\gg \ra \tt$. CP violation in the production process can lead to a definite 
pattern of deviation of $t$ and $\overline{t}$ polarizations from the 
predictions of the standard model (SM). 
A specific example is the asymmetry between the rate of 
production of $t_L\overline{t}_L$ and $t_R\overline{t}_R$, where $L,R$ denote
helicities. Since top polarization is measured only by studying distributions
of the decay products, it is advantageous to make predictions for these 
distributions, as far as possible, without reference to details of top 
reconstruction. 

The usual procedure is to study either CP-violating asymmetries or correlations
of CP-violating observables to get a handle on the CP-violating parameters
of the underlying theory. For a given situation, correlations of optimal
CP-violating variables correspond to the maximum statistical sensitivity. It
is however convenient sometimes to consider variables which are 
simpler and can
be handled more easily theoretically and experimentally.

CP violation can be studied in $\ee$ collisions as well in the $\gamma\gamma$
collider option. Both these approaches are outlined below.
In either case, it is seen that polarized beams help to increase the 
sensitivity. 

SM does not predict measurable CP violation in the processes discussed here. Hence
observation of a nonzero effect would signal physics beyond the standard model.
Most popular extensions of SM predict electric dipole moment of the top quark to 
be at most $10^{-19}-10^{-18}$ $e$ cm, which are extremely difficult to measure.
As will be seen, methods described here cannot individually reach these
limits, and would have to be taken in conjunction with one 
another to be able to reach such sensitivities.

A review of CP violation in top physics can be found in \cite{atwood}.
\section{CP violation studies in $\ee \rightarrow t\overline{t}$}
CP violation in $\ee\ra\tt$ can mainly arise through the couplings of the top
quark to a virtual photon and a virtual $Z$, which are responsible for
$\tt$ production, and the $tbW$ coupling responsible for the dominant decay 
of the top quark into a $b$ quark and a $W$. The CP-violating couplings of 
a $\tt$ current to $\gamma$ and $Z$ can be written as $ie\Gamma_\mu^j$,
where
\beq
\Gamma_\mu^j\;=
\;\f{c_d^j}{2\,m_t}\,i\g_5\,
(p_t\,-\,p_{\overline{t}})_{\mu},\;\;j\;=\;\g,Z,
\eeq
where
$e\cdg/m_t$ and $e\cdz/m_t$ are the electric and ``weak" dipole couplings.

While these CP-violating couplings may be studied using CP-violating 
correlations among momenta and spins which include the $t$ and $\o{t}$
momenta and spins, it may be much more useful to study 
asymmetries and 
correlations constructed out of the initial $e^+/e^-$ momenta and 
the momenta of the decay products, which are more directly observable. 
In addition, the observables using top spin
depend on the basis chosen \cite{parke,lin}, and would require
reconstruction of the basis which has the maximum sensitivity. In studying 
decay distributions, this problem is avoided.

Correlations of optimal CP-violating observables have been studied by Zhou 
\cite{zhou}. Using purely hadronic or hadronic-leptonic variables, limits
on the dipole moment of the order of $10^{-18}$ $e$ cm are shown to be
possible with $\sqrt{s}=500$ GeV and integrated luminosity of 50 fb$^{-1}$.

Examples of CP-violating asymmetries using single-lepton angular 
distributions
and lepton energy correlations have been studied in \cite{grzad}.
In addition, other CP-violating asymmetries which are functions of lepton energy
have been studied in \cite{pra}. 
Using suitable ranges for
the lepton energy, it is possible to enhance the relative contributions of
CP violation in production and CP violation in decay \cite{pra}.
It was shown \cite{grzad,pra} that charged-lepton angular distributions
are independent of any anomalous $tbW$ couplings. Hence angular asymmetries
of leptons can be used to study CP violation in the \ttbar
production process, without pollution coming from CP violation in top decay.

One-loop QCD corrections can contribute as much as 30\% to $\tt$ production
cross section at $\sqrt{s}=500$ GeV \cite{kod}. It is therefore important to include
these in estimates of sensitivities of CP-violating observables. The effect of 
QCD corrections in the soft-gluon approximation in decay lepton distributions in
$\ee\ra\tt$ was discussed in \cite{sga}. These were incorporated in CP-violating
leptonic angular asymmetries and corresponding limits possible at JLC with 
longitudinal beam polarization were presented in \cite{kek}. These are in the
laboratory frame, do not need accurate detailed top 
energy-momentum reconstruction,
and are insensitive to CP violation 
(or other CP-conserving anomalous effects)
in the $tbW$ vertex. 

Four different asymmetries have been studied in \cite{kek}. In addition to 
two asymmetries $A_{ch}(\theta_0)$ and $A_{fb}(\theta_0)$ with the 
azimuthal angles integrated over, which are simple modifications of the
asymmetries defined in \cite{pou},
there are two others, which we call $A_{ud}(\theta_0)$ and 
$A_{lr}(\theta_0)$, and which depend on azimuthal
distributions of the lepton. A cut-off $\theta_0$ in the forward and 
backward directions is assumed in the polar angle of the lepton. 
$A_{ud}(\theta_0)$ and $A_{lr}(\theta_0)$ were discussed earlier in 
\cite{chang}, but in the absence of the cut-off $\theta_0$.

The charge asymmetry $A_{ch}(\theta_0)$, which is simply the cross section asymmetry
between $l^+$ and $l^-$ with a cut-off $\theta_0$, is defined by
\beq\label{ach}
A_{ch}(\theta_0)=\frac{
{\displaystyle      \int_{\theta_0}^{\pi-\theta_0}}d\theta_l
{\displaystyle          \left( \frac{d\sigma^+}{d\theta_l}
        -   \frac{d\sigma^-}{d\theta_l}\right)}}
{
{\displaystyle      \int_{\theta_0}^{\pi-\theta_0}}d\theta_l
{\displaystyle          \left( \frac{d\sigma^+}{d\theta_l} +
\frac{d\sigma^-}{d\theta_l}\right)}}.
\eeq

The other asymmetry $A_{fb}$ is the lept\-onic forward-backward asy\-mmetry com\-bined
with charge asy\-mmetry, again with the angles within $\theta_0$ of
the forward and back\-ward directions excluded:
\beq\label{afb}
A_{fb}(\theta_0)= \frac{ {\displaystyle
\int_{\theta_0}^{\frac{\pi}{2}}}d\theta_l {\displaystyle
\left( \frac{d\sigma^+}{d\theta_l} +
\frac{d\sigma^-}{d\theta_l}\right)} {\displaystyle
- \int^{\pi-\theta_0}_{\frac{\pi}{2}}}d\theta_l {\displaystyle
\left( \frac{d\sigma^+}{d\theta_l} +    \frac{d\sigma^-}{d\theta_l}
\right)}}
{
{\displaystyle      \int_{\theta_0}^{\pi-\theta_0}}d\theta_l
{\displaystyle          \left( \frac{d\sigma^+}{d\theta_l} +
\frac{d\sigma^-}{d\theta_l}\right)}}.
\eeq

 The up-down asymmetry is defined by
\beq\label{aud}
A_{ud}(\theta_0)=\f{1}{2\,\sigma (\theta_0)}\int_{\theta_0}^{\pi-\theta_0}
\left[
\f{d\,\sigma^+_{\rm up} } {d\,\theta_l}
-\f{d\,\sigma^+_{\rm down} } {d\,\theta_l}
+
\f{d\,\sigma^-_{\rm up}} {d\,\theta_l}
-\f{d\,\sigma^-_{\rm down} } {d\,\theta_l}
\right] {d\,\theta_l} ,
\eeq
Here up/down refers to
$(p_{l^{\pm}})_y\;\raisebox{-1.0ex}{$\stackrel{\textstyle>}{<}$}\;0,\;
\:(p_{l^{\pm}})_y$ being the $y$
component of $\vec{p}_{l^{\pm}}$ with respect to a coordinate system
chosen in the $e^+\,e^-$ center-of-mass (cm) frame so that the
$z$-axis is along $\vec{p}_e$, and the $y$-axis is along
$\vec{p}_e\,\times\,\vec{p}_t$.  The $t\bar{t}$ production
plane is thus the $xz$ plane.  Thus, ``up" refers to the range $0<\phi_l<\pi$,
and ``down" refers to the range $\pi<\phi_l<2\pi$. 

\begin{figure}[tb]
\epsfysize=5truein
\vskip -1in
\epsffile[72 72 480 720]{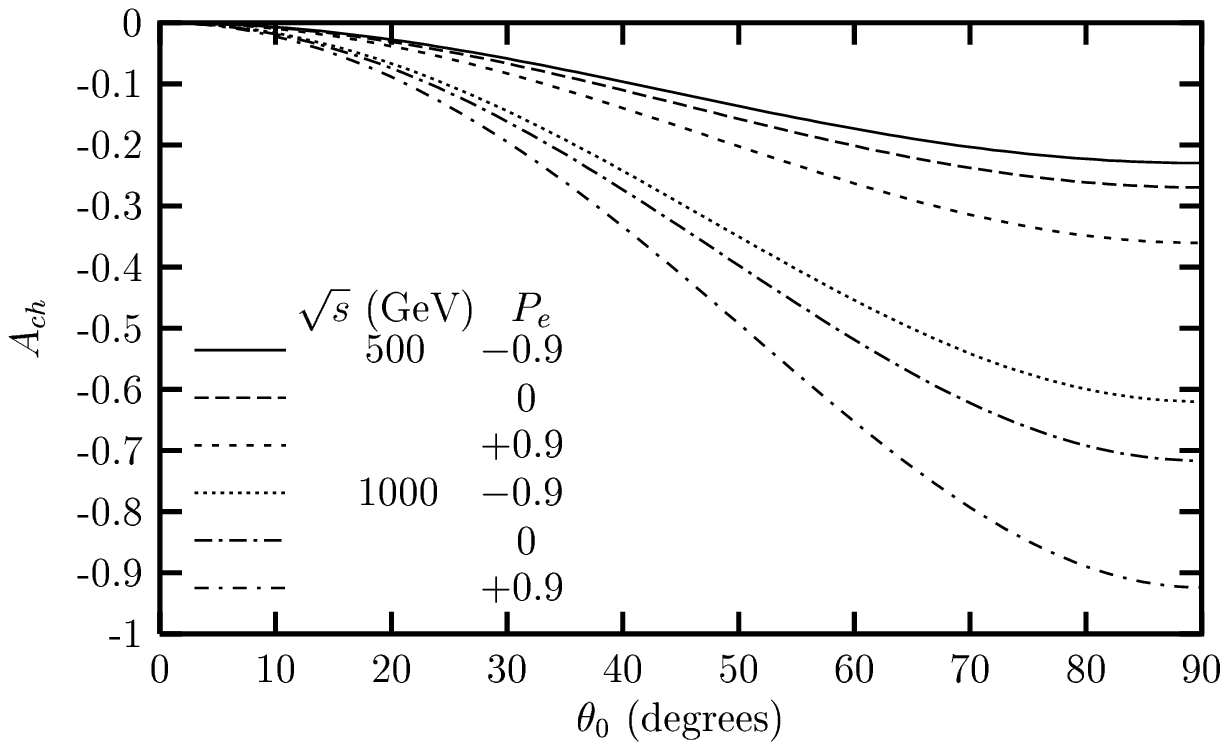}
\epsfysize=5truein
\epsffile[72 72 480 720]{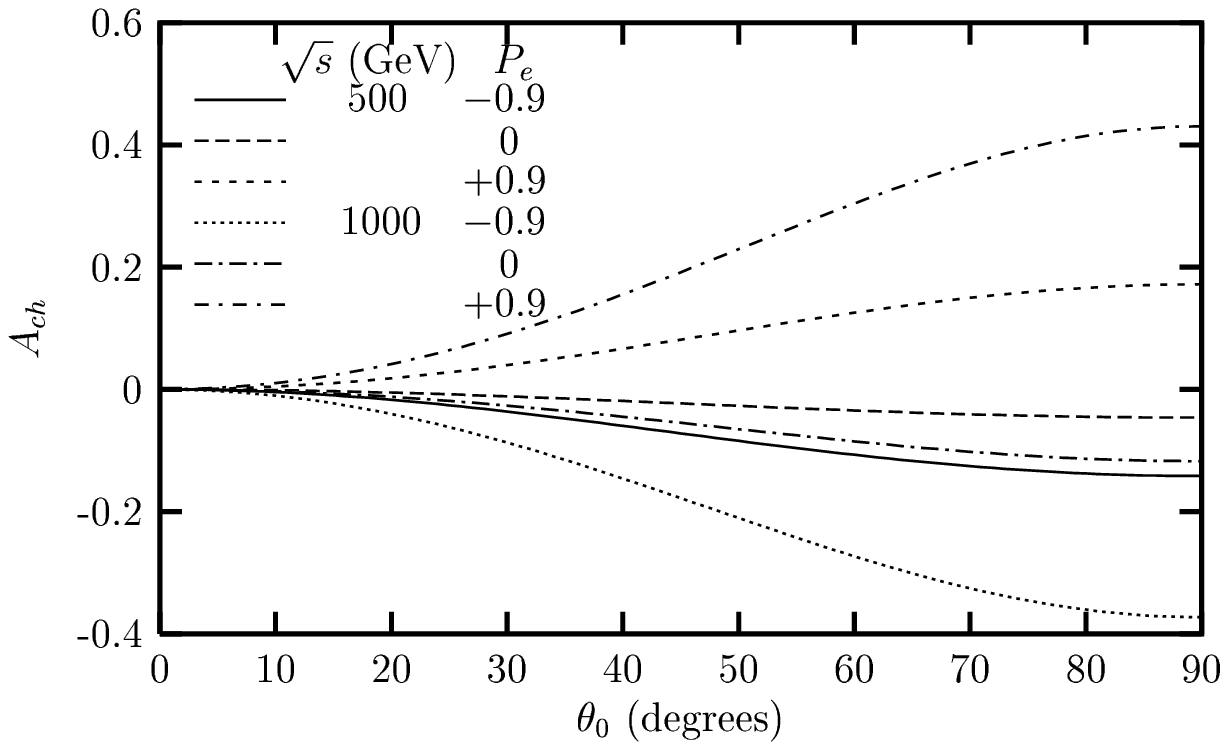}\\[-2in]
\caption{\small
The  asymmetry $A_{ch}$ for Im~$c_d^{\g}=1$, Im~$c_d^Z=0$
(left figure), and for Im~$c_d^{\g}=0$, Im~$c_d^Z=1$ (right figure),
plotted as a function of the cut-off $\theta_0$ on the lepton polar angle
in the forward and backward directions for $e^-$ beam longitudinal
polarizations $P_e=-0.9,0,+0.9$ and for values of total cm energy
$\sqrt{s}= 500$
GeV and $\sqrt{s}=1000$ GeV.}\label{ach.fig}
\epsfysize=5truein
\vskip -1in
\epsffile[72 72 480 720]{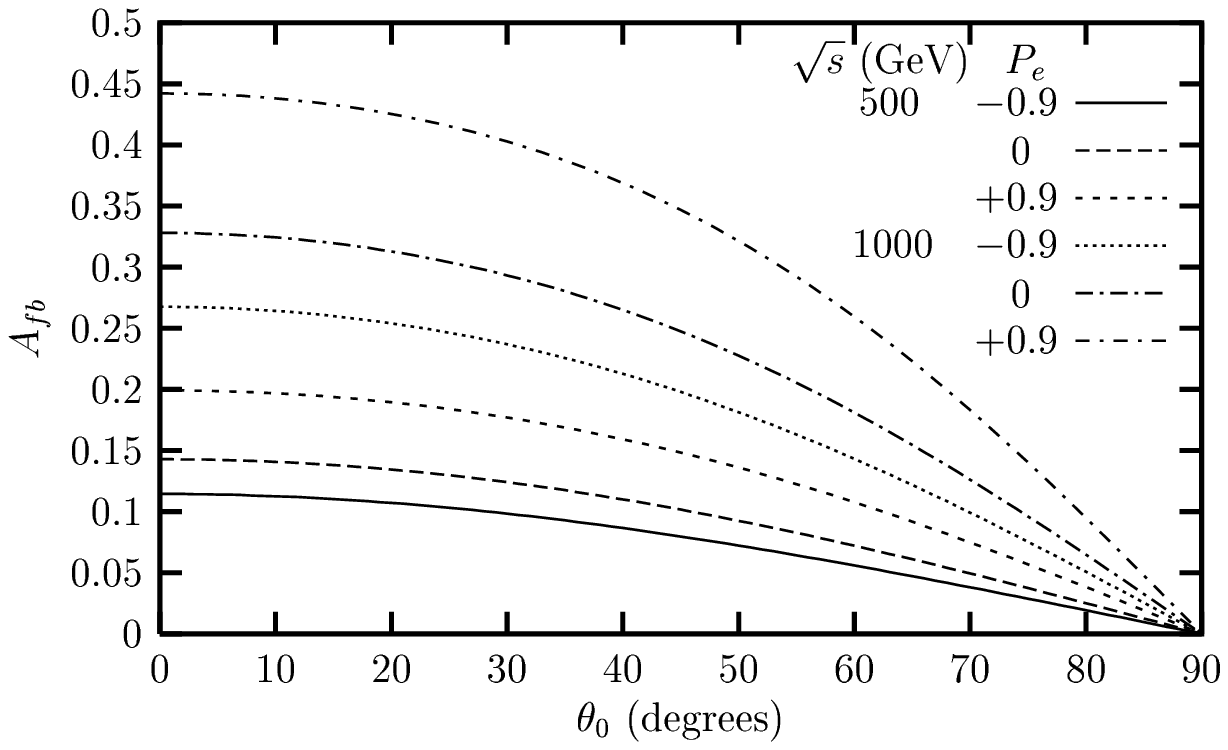}
\epsfysize=5truein
\epsffile[72 72 480 720]{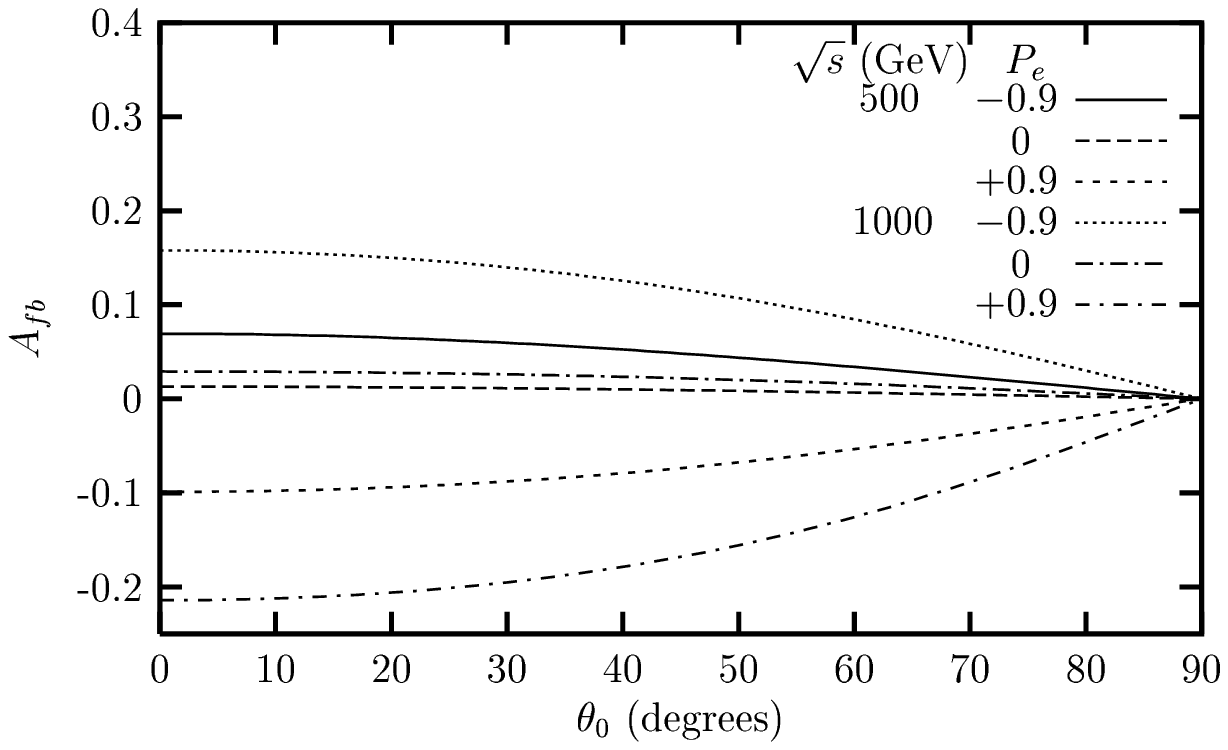}\\[-2in]
\caption{\small
The asymmetry $A_{fb}$ for Im~$c_d^{\g}=1$, Im~$c_d^Z=0$
(left figure), and for  Im~$c_d^{\g}=0$, Im~$c_d^Z=1$ (right figure),
plotted as a function of the cut-off $\theta_0$ on the lepton polar angle
in the forward and backward directions for $e^-$ beam longitudinal
polarizations $P_e=-0.9,0,+0.9$ and for values of total cm energy
$\sqrt{s}= 500$
GeV and $\sqrt{s}=1000$ GeV.}\label{afb.fig}
\end{figure}

The left-right asymmetry is defined by
\beq\label{alr}
A_{lr}(\theta_0)=\f{1}{2\,\sigma (\theta_0)}\int_{\theta_0}^{\pi-\theta_0}
\left[
\f{d\,\sigma^+_{\rm left} } {d\,\theta_l}
-\f{d\,\sigma^+_{\rm right} } {d\,\theta_l}
+
\f{d\,\sigma^-_{\rm left}} {d\,\theta_l}
-\f{d\,\sigma^-_{\rm right} } {d\,\theta_l}
\right] {d\,\theta_l} ,
\eeq
Here left/right refers to
$(p_{l^{\pm}})_x\;\raisebox{-1.0ex}{$\stackrel{\textstyle>}{<}$}\;0,\;
\:(p_{l^{\pm}})_x$ being the $x$
component of $\vec{p}_{l^{\pm}}$ with respect to the coordinate system
system defined above.
Thus, ``left" refers to the range $-\pi /2<\phi_l<\pi /2$,
and ``right" refers to the range $\pi /2<\phi_l<3\pi /2$. 

Fig. \ref{ach.fig} shows the asymmetry $A_{ch}$ 
arising when either of the (imaginary parts of) electric and weak dipole
couplings takes the value 1, the other taking the value 0, plotted as a
function of the cut-off $\theta_0$, for the polarized and unpolarized cases,
for two different cm energies. Fig. \ref{afb.fig} is the corresponding figure
for $A_{fb}$. 

\begin{figure}
\epsfysize=5truein
\vskip -1in
\epsffile[72 72 480 720]{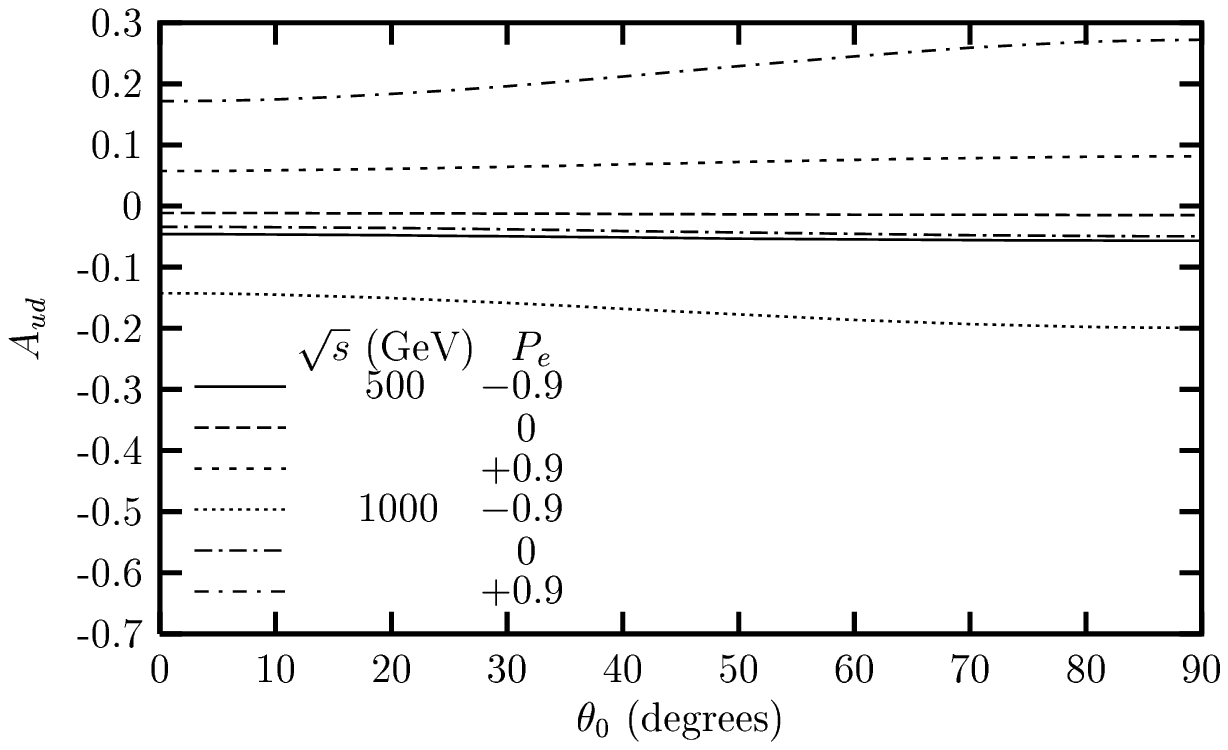}
\epsfysize=5truein
\epsffile[72 72 480 720]{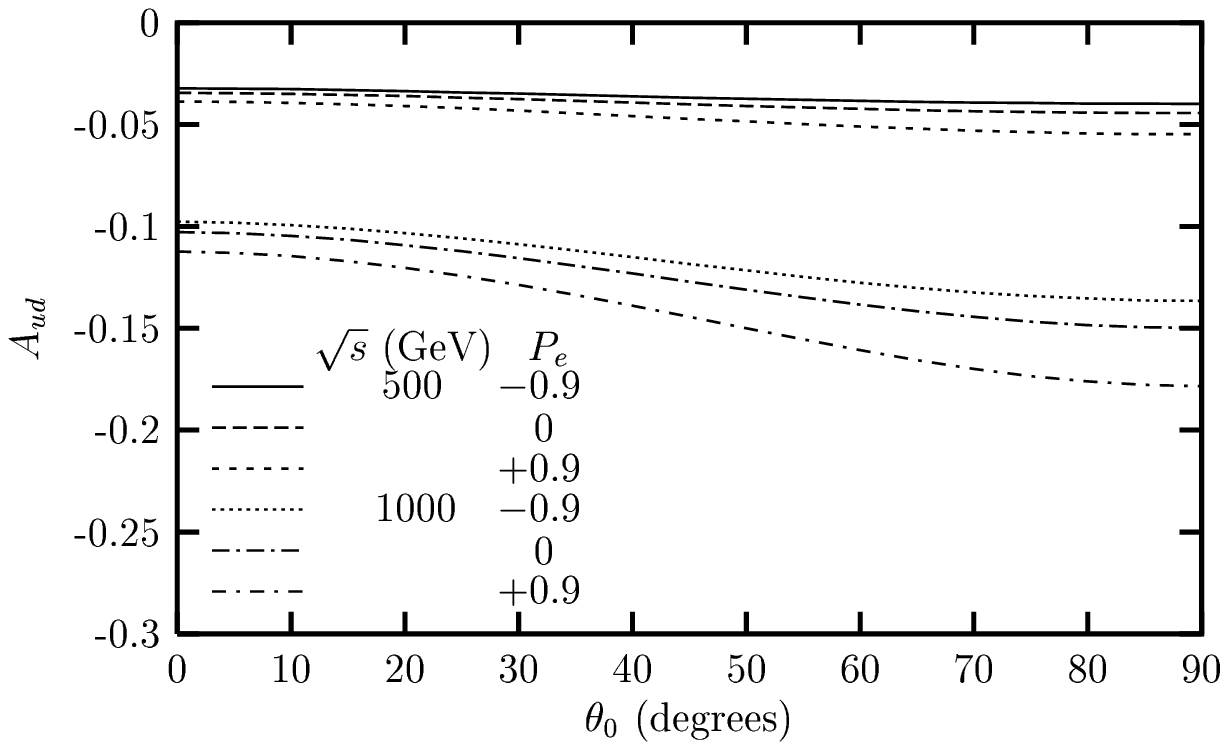}\\[-2in]
\caption{\small
The  asymmetry $A_{ud}$ defined in the text, for Re~$c_d^{\g}=0.1$,
Re~$c_d^Z=0$ (left figure), and for Re~$c_d^{\g}=0$, Re~$c_d^Z=0.1$ (right
figure),
plotted as a function of the cut-off $\theta_0$ on the lepton polar angle
in the forward and backward directions for $e^-$ beam longitudinal
polarizations $P_e=-0.9,0,+0.9$ and for values of total cm energy
$\sqrt{s}= 500$
GeV and $\sqrt{s}=1000$ GeV.}\label{aud.fig}
\epsfysize=5truein
\vskip -1in
\epsffile[72 72 480 720]{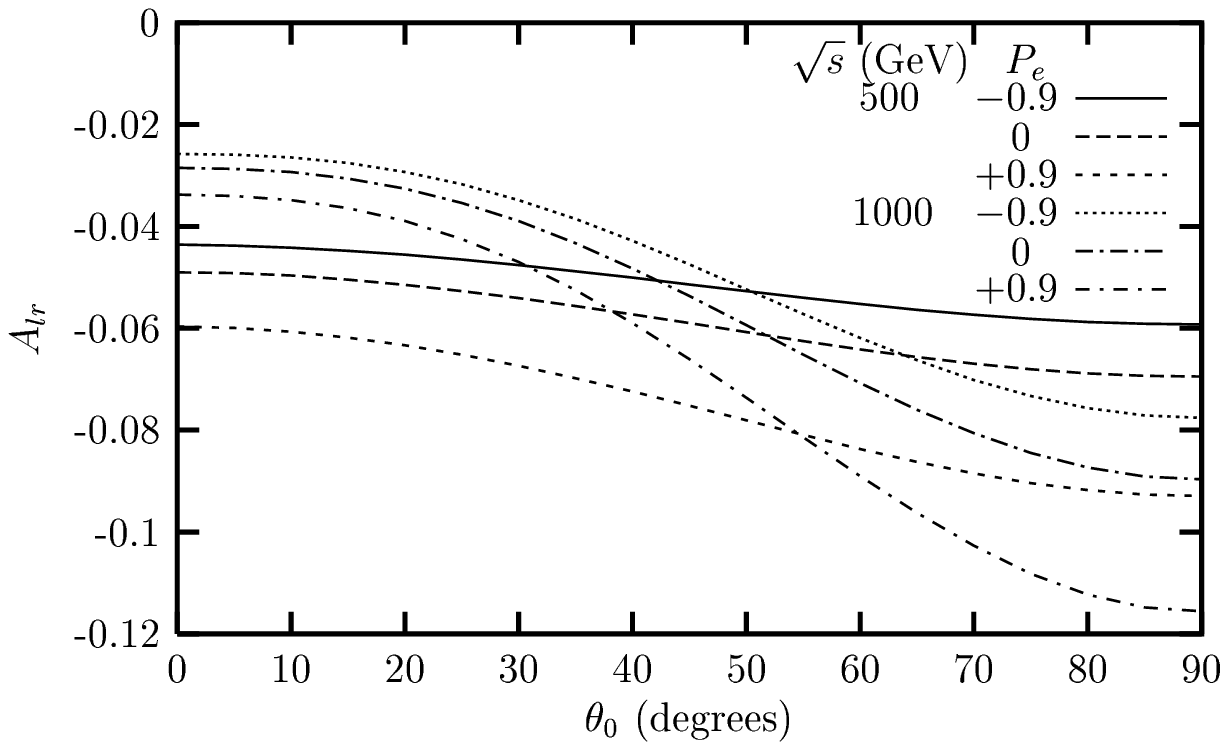}
\epsfysize=5truein
\epsffile[72 72 480 720]{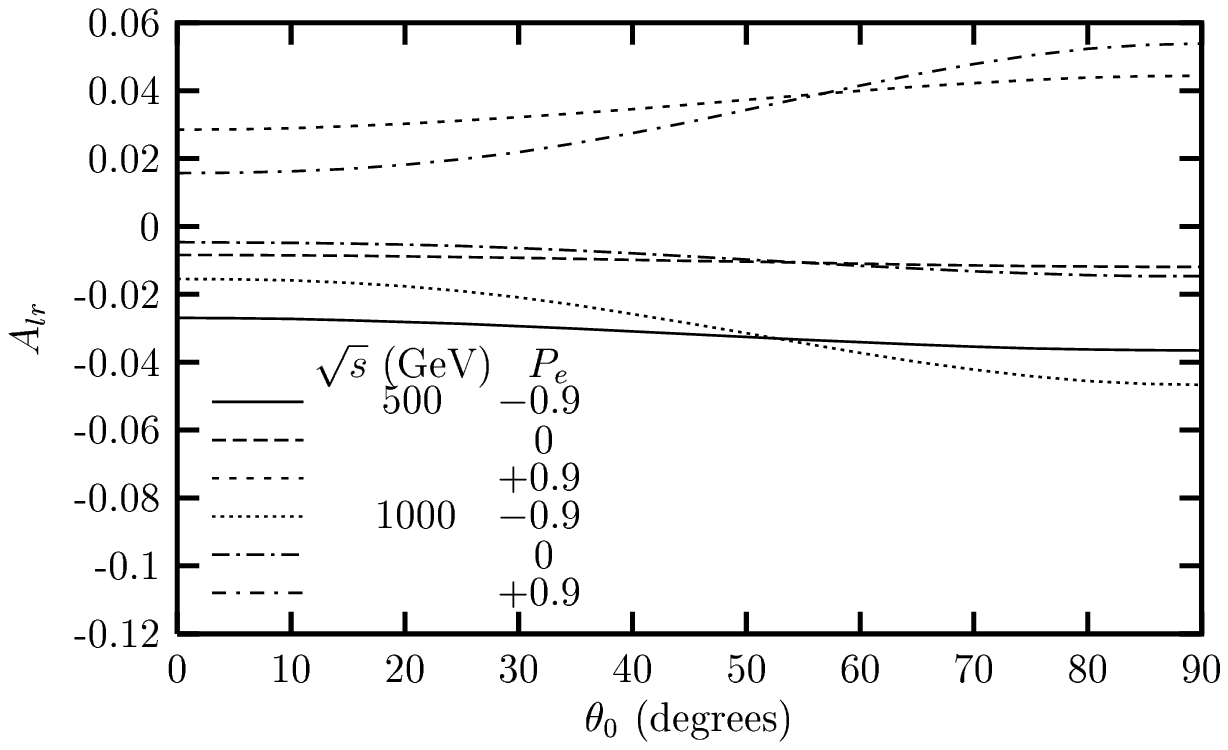}\\[-2in]
\caption{\small
The  asymmetry $A_{lr}$ defined in the text, for Im~$c_d^{\g}=0.1$,
Im~$c_d^Z=0$ (left figure), and for Im~$c_d^{\g}=0$, Im~$c_d^Z=0.1$ (right
figure),
plotted as a function of the cut-off $\theta_0$ on the lepton polar angle
in the forward and backward directions for $e^-$ beam longitudinal
polarizations $P_e=-0.9,0,+0.9$ and for values of total cm energy
$\sqrt{s}= 500$
GeV and $\sqrt{s}=1000$ GeV.}\label{alr.fig}
\end{figure}

Similarly, the asymmetries $A_{ud}$ from eq.~(\ref{aud}) and $A_{lr}$ from
eq.~(\ref{alr}), which depend respectively on the real and imaginary parts of
$c_d^{\g,Z}$, are shown in Figs. \ref{aud.fig} and \ref{alr.fig}. Again, only
one of the couplings takes a nonzero value, in this case 0.1, 
while the others are vanishing.

\begin{table}[tb]
\begin{center}
\begin{tabular}{|c|c|c|c|c|c|c|c|}
\hline
&&
\multicolumn{3}{|c|}{$A_{ch}$}&\multicolumn{3}{|c|}{$A_{fb}$}\\
\cline{3-8}
$\sqrt{s}$ (GeV) &$P_e$ & $\theta_0$ & Im$c_d^{\gamma}$ &
 Im$c_d^Z$ &$\theta_0$ & Im$c_d^{\gamma}$ & Im$c_d^Z$ \\
\hline
    &  0    &64$^{\circ}$ & $ 0.084$ & $ 0.49$ & $10^{\circ}$ & 0.086 & 0.95\\
500 &$+0.9$ &64$^{\circ}$ & $ 0.081$ & 0.17   & $10^{\circ}$ & 0.075 &$ 0.15$\\
    &$-0.9$ &64$^{\circ}$ & $ 0.083$ & $ 0.14$ & $10^{\circ}$ & 0.093 & 0.15 \\
\hline
    & 0      &64$^{\circ}$ & $ 0.029$ & $ 0.18$ & $10^{\circ}$ &0.032&0.36 \\
1000& $+0.9$ &64$^{\circ}$ & $ 0.028$ & 0.061 &   $10^{\circ}$ &0.028&$ 0.058$\\
    & $-0.9$ &64$^{\circ}$ & $ 0.028$ &$ 0.047$&  $10^{\circ}$ &0.034&$ 0.058$\\
\hline
\end{tabular}
\caption{\small
Individual 90\% CL limits on dipole couplings obtainable from
$A_{ch}$ and
$A_{fb}$ for $\sqrt{s}=500$ GeV with integrated luminosity 200 fb$^{-1}$
and for $\sqrt{s}=1000$ GeV with integrated luminosity 1000 fb$^{-1}$ for
different electron beam polarizations $P_e$. Cut-off $\theta_0$ is chosen to
optimize the sensitivity.}
\end{center}
\end{table}

Tables 1-5
 show the results on the limits obtainable for each of these
possibilities. In all cases, the value of the cut-off $\theta_0$ has been
chosen to get the best sensitivity for that specific item. In case of
$A_{fb}$, the sensitivity is maximum for $\theta_0=0$. In that case, the
cut-off has been arbitrarily chosen to be 10$^\circ$.

\begin{table}
\begin{center}
\begin{tabular}{|c|c|c|c|}
\hline
$\sqrt{s}$ (GeV)  & $\theta_0$ & Im$c_d^{\gamma}$ & Im$c_d^Z$\\
\hline
500 & 40$^{\circ}$ & 0.53 & 4.1 \\
1000& 40$^{\circ}$ & 0.20 & 1.5 \\
\hline
\end{tabular}
\caption{\small
Simultaneous 90\% CL limits on dipole couplings obtainable from
$A_{ch}$ and
$A_{fb}$ for $\sqrt{s}=500$ GeV with integrated luminosity 200 fb$^{-1}$
and for $\sqrt{s}=1000$ GeV with integrated luminosity 1000 fb$^{-1}$ for
unpolarized beams. Cut-off $\theta_0$ is chosen to
optimize the sensitivity.}
\end{center}
\end{table}

In Table 1 are given the 90\% confidence level (CL) limits that can be 
obtained on Im$c_d^{\g}$ and
Im$c_d^Z$, assuming one of them to be nonzero, the other taken to be vanishing.
The limit is defined as the value of Im$\cdg$ or Im$\cdz$ for which the
corresponding asymmetry $A_{ch}$ or $A_{fb}$ becomes equal to 
$1.64/\sqrt{N}$, where $N$ is the total number of events.

\begin{table}[b]
\begin{center}
\begin{tabular}{|c|c|c|c|c|c|c|}
\hline
&
\multicolumn{3}{|c|}{$A_{ch}$}&\multicolumn{3}{|c|}{$A_{fb}$}\\
\cline{2-7}
$\sqrt{s}$ (GeV)&  $\theta_0$ & Im$c_d^{\gamma}$ &
 Im$c_d^Z$ &$\theta_0$ & Im$c_d^{\gamma}$ & Im$c_d^Z$ \\
\hline
500 & 64$^{\circ}$ & 0.11 & 0.20 & $10^{\circ}$ & 0.11 & 0.20 \\
1000& 64$^{\circ}$ & 0.037& 0.069 &$10^{\circ}$ & 0.040 & 0.076 \\
\hline
\end{tabular}
\caption{\small Simultaneous limits on dipole couplings combining data from
polarizations $P_e=0.9$ and $P_e=-0.9$, using separately $A_{ch}$ and
$A_{fb}$ for $\sqrt{s}=500$ GeV with integrated luminosity 200 fb$^{-1}$
and for $\sqrt{s}=1000$ GeV with integrated luminosity 1000 fb$^{-1}$.
Cut-off $\theta_0$ is chosen to optimize the sensitivity.}
\end{center}
\end{table}

Table 2 shows possible 90\% CL limits for the unpolarized case, when results
from $A_{ch}$ and $A_{fb}$ are combined.
The idea is that each asymmetry measures a different linear
combination of Im$c_d^{\g}$ and
Im$c_d^Z$. So a null result for the two asymmetries will correspond to two
different bands of regions allowed at 90\% CL in the space of Im$c_d^{\g}$ and
Im$c_d^Z$. The overlapping region of the two bands leads to the limits given in
Table 2. In this case, for 90\% CL, the asymmetry is required to be
$2.15/\sqrt{N}$, corresponding to two degrees of freedom. Incidentally, the 
same procedure followed for $P_e=\pm 0.9$ gives much worse limits.

Similarly, using one of the two asymmetries, but two different polarizations of
the electron beam, one can get two bands in the parameter plane, which give
simultaneous limits on the dipole couplings. The results for electron
polarizations $P_e=\pm 0.9$ are given in Table 3 for each of the asymmetries
$A_{ch}$ and $A_{fb}$.

\begin{table}[t]
\begin{center}
\begin{tabular}{|c|c|c|c|c|c|c|c|}
\hline
&&
\multicolumn{3}{|c|}{$A_{ud}$}&\multicolumn{3}{|c|}{$A_{lr}$}\\
\cline{3-8}
$\sqrt{s}$ (GeV) &$P_e$ & $\theta_0$ & Re$c_d^{\gamma}$ &
 Re$c_d^Z$ &$\theta_0$ & Im$c_d^{\gamma}$ & Im$c_d^Z$ \\
\hline
    &  0    &25$^{\circ}$ & $ 0.10$ & $ 0.034$ & $30^{\circ}$ & 0.024 & 0.14\\
500 &$+0.9$ &30$^{\circ}$ & $ 0.025$ & 0.037  & $35^{\circ}$ & 0.024 & 0.050\\
    &$-0.9$ &25$^{\circ}$ & $ 0.022$ & 0.032 & $30^{\circ}$ & 0.023 &0.038\\
\hline
    & 0      &30$^{\circ}$ & $ 0.029$ & $ 0.0096$ & $60^{\circ}$ &0.021&0.13 \\
1000& $+0.9$ &35$^{\circ}$ &  0.0068 & 0.010 &   $60^{\circ}$ &0.021&$ 0.045$\\
    & $-0.9$ &30$^{\circ}$ &  0.0061& 0.0089 &  $60^{\circ}$ &0.021&$ 0.035$\\
\hline
\end{tabular}
\caption{\small Individual 90\% CL limits on dipole couplings obtainable from
$A_{ud}$ and
$A_{lr}$ for $\sqrt{s}=500$ GeV with integrated luminosity 200 fb$^{-1}$
and for $\sqrt{s}=1000$ GeV with integrated luminosity 1000 fb$^{-1}$ for
different electron beam polarizations $P_e$. Cut-off $\theta_0$ is chosen to
optimize the sensitivity.}
\end{center}
\end{table}

\begin{table}[t]\label{t5}
\begin{center}
\begin{tabular}{|c|c|c|c|c|c|c|}
\hline
&
\multicolumn{3}{|c|}{$A_{ud}$}&\multicolumn{3}{|c|}{$A_{lr}$}\\
\cline{2-7}
$\sqrt{s}$ (GeV)&  $\theta_0$ & Re$c_d^{\gamma}$ &
 Re$c_d^Z$ &$\theta_0$ & Im$c_d^{\gamma}$ & Im$c_d^Z$ \\
\hline
500 & 25$^{\circ}$ & 0.031& 0.045& $35^{\circ}$ & 0.031& 0.056\\
1000& 30$^{\circ}$ & 0.0085& 0.013 &$60^{\circ}$ & 0.028 & 0.052 \\
\hline
\end{tabular}
\caption{\small Simultaneous limits on dipole couplings combining data from
polarizations $P_e=0.9$ and $P_e=-0.9$, using separately $A_{ud}$ and
$A_{lr}$ for $\sqrt{s}=500$ GeV with integrated luminosity 200 fb$^{-1}$
and for $\sqrt{s}=1000$ GeV with integrated luminosity 1000 fb$^{-1}$. Cut-off
$\theta_0$ is chosen to optimize the sensitivity.}
\end{center}
\end{table}

Table 4 lists the 90\% CL limits which may be obtained on the real and
imaginary parts of the dipole couplings using $A_{ud}$ and $A_{lr}$, assuming
one of the couplings to be nonzero at a time.

Table 5 shows simultaneous limits on Re$c_d^{\g}$ and Re$c_d^Z$ obtainable from 
combining the data on $A_{ud}$ for $P_e=+0.9$ and $P_e=-0.9$, and similarly,
limits on Im$c_d^{\g}$ and Im$c_d^Z$ from data on $A_{lr}$ for the two
polarizations.

\section{CP violation studies in $\gamma\gamma \rightarrow t\overline{t}$}

CP-violating dipole couplings of the top quark to photons can be studied at 
$\gamma\gamma$ colliders. The advantage over the study using $\ee$ collisions
is that the electric dipole moment is obtained independent of the weak dipole
coupling to $Z$. Several proposals exist for the study of CP violation at $\g\g$
colliders.

Ma et al. \cite{ma} have discussed the 
CP-violating couplings of 
neutral Higgs in the context of a two-Higgs doublet model. They studied the 
CP-violating asymmetries 
\begin{equation}
\xi_{\rm CP} = \frac{\sigma_{t_L\bar{t}_L} - \sigma_{t_R\bar{t}_R}}
		{\sigma_{t\bar{t}}},
\end{equation}
for the case of unpolarized photon beams, and
\begin{equation}
\xi_{\rm CP,1 } = \frac{\sigma^{++}_{t_L\bar{t}_L} 
		- \sigma^{--}_{t_R\bar{t}_R}}
                 {\sigma^{++}_{t_L\bar{t}_L} + \sigma^{--}_{t_R\bar{t}_R}}
\end{equation}  
and
\begin{equation}
\xi_{\rm CP,2 } = \frac{\sigma^{++}_{t_R\bar{t}_R} 
		- \sigma^{--}_{t_L\bar{t}_L}}
                 {\sigma^{++}_{t_R\bar{t}_R} + \sigma^{--}_{t_L\bar{t}_L}}
\end{equation}  
for the case of circularly polarized photon beams, where the superscripts on
$\sigma$ denote the signs of the photon helicities, and the subscripts $L$ and
and $R$ denote left- and right-handed polarizations for the quarks. They found
that asymmetries of the order of 10$^{-4}$ -- 10$^{-3}$ can get enhanced to 
the level of a few percent in the presence of beam polarization, for reasonable
values of the model parameters.
Similar CP-violating observables have been identified in the MSSM by M.-L. Zhou et
al.\cite{mlzhou}.

Choi and Hagiwara \cite{choiedm} and Baek et al. 
\cite{baek} have proposed the study of the number asymmetry of top quarks with 
linearly polarized photon beams, and found that a limit of about 10$^{-17}$
$e$ cm can be put on the electric dipole moment (edm) of the top quark with an 
integrated $\ee$ luminosity of 20 fb$^{-1}$ for $\sqrt{s}=500$ GeV. 

\begin{table}[hb]
\begin{center}
\begin{tabular}{|r|r|r|r|r|r|r|r|c|}
\hline
&&&&&\multicolumn{2}{|c|}{ }&
\multicolumn{2}{|c|}{Limits on Im $d_t$ }\\
&&&&&\multicolumn{2}{|c|}{ Asymmetries}&\multicolumn{2}{|c|}{(in
$10^{-16}\;e\,cm$)}\\
&&&&&\multicolumn{2}{|c|}{ }&
\multicolumn{2}{|c|}{from}\\[2mm]
$\lambda_e^1$&$\lambda_e^2$&$\lambda_l^1$&$\lambda_l^2$&
N&$A_{ch}$&$A_{fb}$&$|A_{ch}|$&$|A_{fb}|$\\[2mm]
\hline
$-$0.5& $-$0.5&$-$1&$-$1&55  & $-$0.019 & 0 &3.22&\\
$-$0.5& $-$0.5&   1&$-$1&215 & $-$0.025 &$-$0.129 &1.28&0.25\\
$-$0.5& $-$0.5&   1&   1&631 & $-$0.035 & 0 &0.54&\\
   0.5& $-$0.5&$-$1&$-$1&63  & $-$0.024 & 0.013 &2.49&4.58\\
   0.5& $-$0.5&   1&$-$1&23  & $-$0.021 &$-$0.080&4.53&1.20\\
   0.5& $-$0.5&$-$1&   1&163 & $-$0.021 & 0.033&1.72&1.11\\[2mm]
\multicolumn{4}{|c|}{Unpolarized}&179&$-$0.024&0&1.44&\\
\hline
\end{tabular}
\caption{\small Asymmetries and corresponding 90\% C.L. limits obtained on Im $d_t$ 
for various combinations of initial-beam helicities. The initial electron beam
energy of $E_b=250$ GeV, laser beam energy of $\omega_0=1.24$ eV and cut-off 
angle $\theta_0=30^{\circ}$ are assumed. $N$ is the total number of events, 
and the asymmetries are for a value Im $d_t=1/(2m_t)$.}\label{hel}
\end{center}
\end{table}

\begin{table}[ht]
\begin{center}
\begin{tabular}{|c|r|c|c|c|c|}
\hline
&&\multicolumn{2}{|c|}{}&
\multicolumn{2}{|c|}{ Limits on Im $d_t$ }\\
&&\multicolumn{2}{|c|}{Asymmetries}&
\multicolumn{2}{|c|}{(in $10^{-16}\;e\,cm)$}\\
&&\multicolumn{2}{|c|}{}&
\multicolumn{2}{|c|}{
from}\\
\multicolumn{1}{|c|}{$E_b$ (GeV)}&N&$A_{ch}$&$A_{fb}$&$|A_{ch}|$&$|A_{fb}|$\\[1mm]
\hline
250&  215 &$-$0.025 & 0.129&1.282&  0.247\\
500&  1229&$-$0.167 & 0.420& 0.080& 0.032\\
750&  1033&$-$0.223 & 0.347& 0.065& 0.042\\
1000& 850 &$-$0.227 & 0.244& 0.071& 0.066\\
\hline
\end{tabular}
\caption{\small Number of events, asymmetries and limits on Im $d_t$, as in previous 
table, as a function of the beam electron beam energy $E_b$.}\label{ebeam} 
\end{center}
\end{table}

Asymmetries of charged leptons from 
top decay in $\gg \ra \tt$ with longitudinally polarized photons have been
studied in \cite{pp,ppconf}. These 
asymmetries do not need full reconstruction of the top or anti-top. Generalizations
of the asymmetries $A_{ch}$ and $A_{fb}$ described in the previous section 
(eqs. (\ref{ach}) and (\ref{afb})) can be used to measure the electric dipole moment.
The total number of events, the asymmetries and the 90\% C.L. limits on the dipole 
moment at shown for various helicity combinations of the initial beams 
in Table \ref{hel}. Here a cut-off angle $\theta_0$ of $30^{\circ}$ 
is assumed, and the beam
energy is taken to be 250 GeV. The geometric luminosity is assumed to be 
20 fb$^{-1}$.
The laser beam energy is assumed to be 1.24 eV. Table \ref{ebeam} shows 
the variation of various results with beam energy.  Limits 
at the 90\% confidence level 
of the order of $2\times 10^{-17}$ $e$ cm on (the imaginary part of) 
the top edm can be obtained with
an $\ee$  luminosity of 20 fb$^{-1}$ and cm energy $\sqrt{s}=500$ GeV and 
suitable choice of electron beam and laser photon polarizations. The
limit can be improved by a factor of 8 by going to $\sqrt{s}=1000$ GeV.

It should be emphasized that the method relies on direct observation of
lepton asymmetries rather than top polarization asymmetries, and hence 
does not depend heavily on the accuracy of top reconstruction.

We have  also considered \cite{pp2,ppconf} the asymmetries discussed
in \cite{pp} to study the 
simultaneous presence of the top edm and an effective CP-violating 
$Z\gamma\gamma$ coupling. By using two different decay-lepton asymmetries,
the top edm coupling and the $Z\gamma\gamma$ coupling can be studied in 
a model-independent manner.

Asakawa et al. \cite{asa} study the possibility of determining completely the 
effective couplings of a neutral Higgs scalar to two photons and to 
$t\overline{t}$ when CP is violated. They study the effects of a neutral Higgs 
boson without definite CP parity in the process $\gamma\gamma \rightarrow 
t\overline{t}$ around the pole of the Higgs boson mass. Near the resonance,
interference between Higgs exchange and the continuum SM amplitude can be
sizeable. This can permit a measurement in a model-independent way of 6 coupling
constant combinations by studying cross sections with initial beam polarizations
and/or final $t$, $\overline{t}$ polarizations. Using general (circular
as well as linear) polarizations for
the two photons, and different longitudinal polarizations for the $t$ 
and $\overline{t}$, in all 22 combinations 
could be measured, which could be used to determine the 6 parameters of 
the theory. Of these, half are CP-odd, and the remaining are CP-even.
They also consider a specific example of MSSM, where CP-odd measurements are 
sensitive to the case of low $\tan\beta$.

To enable the use of observables directly measureable, we are in the process 
of evaluating angular and energy distributions of decay leptons in the 
laboratory frame \cite{ritesh}. Distributions are calculated for arbitrary
photon polarizations and folded with the photon spectra arising from 
Compton scattering of polarized photons off longitudinally 
polarized electrons
or positrons. Work is under progress to estimate CP-violating asymmetries 
and to examine the sensitivity of these to the CP-violating couplings.

\section{Conclusions}

The processes \eett~ and \ggtt~ can be useful probes of CP
violating electric and weak dipole couplings of the top quark to $\gamma$ and
$Z$, as also of CP-violating couplings to neutral Higgs.

Polarization can play a useful role in enhancing the sensitivity, as well as
providing an additional parameter to tune for simultaneous model-independent 
determination of more than one parameter.

Results were presented for angular asymmetries of decay leptons in the
laboratory frame, including ${\cal O}(\alpha_s)$ QCD corrections, for the
$\ee$ option.
Only CP violation in production contributes to these asymmetries. 
Simultaneous model-independent 90\% CL limits of order $10^{-17}$ $e$ cm 
can be placed on electric and weak dipole moments of the top quark for 
$\sqrt{s}= 500$ GeV and integrated luminosity of 200 fb$^{-1}$, with electron beam
polarization of $\pm 0.9$.
The limit can be improved by a large factor by going to $\sqrt{s}= 1000$ GeV
and integrated luminosity 1000 fb$^{-1}$.

Using the $\g\g$ option the electric dipole coupling can be measured
independent of the weak dipole coupling. Linear or circular polarization of
photons can be used as an extra handle to isolate useful CP-odd asymmetries.
Again asymmetries of decay-lepton angles in the laboratory frame were shown to
be useful in studying CP violation. Limits on the top edm are of order of a few
times $10^{-17}$ $e$ cm for the 500 GeV option, and somewhat better for higher
$\sqrt{s}$.

The $\g\g$ option may also be used to study CP-violating $\g\g H$ couplings. 
Initial photon polarization combinations provide a number of asymmetries which
could be measured. Decay lepton angular distributions in the laboratory frame
could provide direct measurement of the CP-violating couplings.

It is clear that to reach sensitivities which can probe predictions of
various extensions of SM which predict top electric dipole moment in the 
range of $10^{-19} - 10^{-18}$ $e$ cm, measurements from different 
asymmetries and from several decay 
channels, including hadronic ones,  may have to be combined.
Most important of all, future studies should concentrate on putting in
realistic kinematic cuts against backgrounds and realistic detector
efficiencies.


\thebibliography{cc}
\bibitem{atwood} D. Atwood, S. Bar-Shalom, G. Eilam  and A. Soni,
Phys. Rep. {\bf 347}, 1 (2001). 
\bibitem{parke} S. Parke and Y. Shadmi, Phys. Lett. B {\bf 387}, 199 (1996);
Y. Kiyo, J. Kodaira, K. Morii, T. Nasuno and S. Parke, Nucl. Phys. Proc. 
Suppl. {\bf 89}, 37 (2000); Y. Kiyo, J. Kodaira and K. Morii, Eur. Phys. J. C
{\bf 18}, 327 (2000).
\bibitem{lin} Z.-H. Lin, T. Han, T. Huang, J.-X. Wang and X. Zhang, 
hep-ph/0106344.
\bibitem{zhou} H.-Y. Zhou, Phys. Lett. {\bf 439}, 393 (1998).
\bibitem{grzad} B. Grzadkowski and Z. Hioki, Phys. Lett. B {\bf 476}, 87 (2000);
Nucl. Phys. B {\bf 585}, 3 (2000).
\bibitem{pra} S.D. Rindani, Pramana J. Phys. {\bf 54}, 791 (2000).
\bibitem{kod} J. Kodaira , T. Nasuno and S. Parke, Phys. Rev. D {\bf 59}, 
014023 (1999).
\bibitem{sga} S.D. Rindani, Phys. Lett. B {\bf 503}, 292 (2001).
\bibitem{kek} S.D. Rindani, hep-ph/0105318, to be published in the proceedings
of the Theory Meeting on Linear Colliders, KEK, Japan, March 16-23, 2001;
S.D. Rindani (in preparation).
\bibitem{pou} P. Poulose and S.D. Rindani,  Phys. Rev. D {\bf 54},
4326 (1996); {\bf 61}, 119901 (2000) (E); Phys. Lett. B {\bf 383}, 212 (1996).
\bibitem{ma} W.-G. Ma {\it et al.}, Commun. Theor. Phys. {\bf 26}, 
455 (1996); {\bf  27}, 101 (1997).
\bibitem{chang} D. Chang and W.-Y. Keung, Nucl. Phys. B {\bf 408}, 286 
(1993), Nucl. Phys. B {\bf 429}, 255 (1994) (E); P. Poulose and S.D. 
Rindani, Phys. Lett. B {\bf 349}, 379 (1995).  
\bibitem{mlzhou} M.-L. Zhou {\it et al.}, J. Phys. G {\bf 25}, 27 (1999). 
\bibitem{choiedm} S.Y. Choi and K. Hagiwara, Phys. Lett. B {\bf 359}, 369 (1995).
\bibitem{baek} M.S. Baek, S.Y. Choi and C.S. Kim, Phys. Rev. D {\bf 56}, 6835 (1997).
\bibitem{pp} P. Poulose and S.D. Rindani, Phys. Rev. D {\bf 57}, 5444 (1998);
{\bf 61}, 119902 (2000) (E).
\bibitem{ppconf} P. Poulose, Nucl. Instr. Meth. A {\bf 472}, 195 (2001).
\bibitem{pp2} P. Poulose and S.D. Rindani, Phys. Lett. B {\bf 452}, 347 (1999).
\bibitem{asa} E. Asakawa, S.Y. Choi, K. Hagiwara and J.S. Lee, Phys. Rev. D 
{\bf 62}, 115005 (2000).
\bibitem{ritesh} R.M. Godbole, S.D. Rindani and R.K. Singh, work in progress.
\end{document}